\def\G1915{GRS $1915$+$105$}
\def\X1550{XTE J$1550$--$564$}
\def\J1655{GRO J$1655$--$40$}

\def\aql{Aql~X-1}
\def\chis{$\chi^2_\nu$}

\def\ergcms{erg cm$^{-2}$ s$^{-1}$ }
\def\nh{$N_{\mathrm H}$}

\def\integ{\textit{INTEGRAL}}
\def\rxte{\textit{RXTE}}

\documentclass[]{aa}
\usepackage{epsf,epsfig}
\usepackage{color}
\definecolor{red}{rgb}{0.7,0,0}
\definecolor{blue}{rgb}{0,0,0.7}

\begin{document}

\titlerunning{The 2005 hard state outburst of Aql X-1}
\authorrunning{Rodriguez, Shaw \& Corbel}
\title{The faint 2005  hard state outburst of Aquila X-1 seen by INTEGRAL and RXTE}

\author{J. Rodriguez\inst{1}, S.E. Shaw\inst{2,3} \& S. Corbel\inst{1}}

\institute{AIM - Unit\'e Mixte de Recherche CEA - CNRS - Universit\'e Paris VII
 - UMR 7158, CEA Saclay, Service d'Astrophysique, F-91191 Gif sur Yvette, France \and School of 
Physics and Astronomy, University of Southampton, 
SO 17 1BJ, UK \and INTEGRAL Science Data Centre, 16 Chemin d'Ecogia, CH-1290 Versoix, Switzerland}

\date{received ; accepted}

\abstract{We report on the spectral analysis of \rxte\ and \integ\ data of 
the 2005 April outburst of the transient Atoll source \aql. 
Although this outburst is one of the faintest ever detected
for this source in the soft 
X-rays (\rxte/ASM), one of our \integ\ observations, taken close to the soft 
X-ray peak, shows that the source   flux was quite high, with a 
20-200 keV flux of $2.05\times10^{-9}$ \ergcms. On this occasion we detect the source up to 
150 keV for the first time. We compare and discuss the similarity of the source behavior 
with that of black hole transients especially XTE J1550$-$564.
}

\maketitle

\section{Introduction}
\aql\ is a transient X-ray binary which undergoes outbursts about once per year.
It is a Low Mass X-ray binary composed of a Roche-lobe filling late type K star
orbiting a weakly magnetized neutron star at a distance 4.5-6 kpc 
(Chevalier et al. 1999; Rutledge et al. 2001). Along its outbursts it has been observed 
to transit through several spectral states, either characterized by a hard power law tail, 
and or the presence of a soft X-ray component usually modeled by an optically thick 
accretion disk 
(e.g. Maitra \& Bailyn 2004). The  short recurrence time around 1 year (and its evolution 
along  the outbursts makes it an ideal target to study the accretion processes around 
compact objects. \\
\indent Apart from very bright outbursts, \aql\ is sometimes seen to undergo ``low luminosity'' 
(in the soft X-rays) hard state outbursts (a.k.a mini-outburst). Although these were already 
reported with BATSE (Harmon et al. 1996), it is remarkable that the two last outbursts were of 
this type (e.g. Rodriguez et al. 2004, Maitra and Bailyn 2005b, for the 2004 
outburst). On April 1st 2005 the source 
was detected during an \integ\ Galactic Plane Scan (Grebenev et al. 2005). This detection 
followed the apparent reactivation of the source in optical (Maitra \& Bailyn 2005a), suggesting a new
outburst was beginning. The source was again in the field of view of our \integ\ monitoring
campaign of GRS~1915+105 (PI Rodriguez) 10 days and $\sim 40$ days later. Very recently, 
Rodriguez \& Shaw (2005) reported renewed activity of \aql\ in the hard X-rays with INTEGRAL in 
the end of November 2005.\\
\indent We report here the spectral analysis of the \integ\ and \rxte\ observations. We start 
by giving the data reduction procedures in the next section, and present the results of our 
observation in section 3. We briefly discuss our findings in the last part of the paper.
 
\section{Observations and data reduction}
\begin{figure}
\centering
\epsfig{file=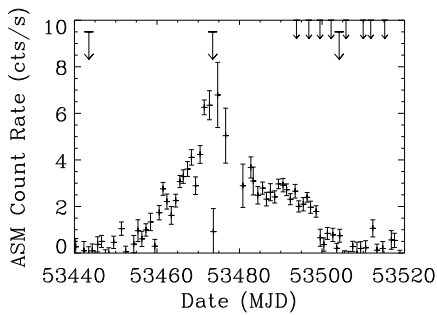,width=6.5cm}
\caption{1.2--12 keV \rxte/ASM light curve of \aql\ during the 2005 outburst. The small arrows indicate
the moment of RXTE dedicated pointings (PCA+HEXTE), and the long ones the dates 
of our \integ\ observations.}
\label{fig:lc}
\end{figure}

Fig. \ref{fig:lc} shows the \rxte/ASM light curve during the 2005 outburst. The individual 
\integ\ pointings are indicated. The journal of the observations is reported in 
Table \ref{tab:loginteg}.

\begin{table}
\caption{Journal of the \integ\ and \rxte\ observations presented in this paper. 
$^\star$Effective exposure
of the \aql\ spectra, obtained after correction for gti and instrumental deadtime.}
\begin{tabular}{ccccc}
\hline
\multicolumn{4}{c}{\integ\ }\\
\hline
Rev. \# & Start & Stop  &  Effective Exposure$^\star$ \\
          & MJD & MJD     & ($\times 10^3$ s) \\ 
\hline
\hline

 295 & 53442.95 & 53444.15 & {\textit{Not detected}}\\
 305 & 53472.87 & 53474.14 & 77\\
 315 & 53503.54 & 53504.85 & 51\\
\hline
\multicolumn{4}{c}{\rxte\ }\\
\hline
Obs. \# & ObsId& Start & PCA Good Time  \\
       &   & MJD & (s) \\ 
\hline
\hline
1 & 91414-01-01-00 & 53493.77    &  3216 \\
2 & 91414-01-02-00 & 53496.79    &   2464\\
3 & 91414-01-02-01 & 53499.53      & 3200 \\
4 & 91414-01-02-02 & 53502.24       & 2208\\
5 & 91414-01-03-00 & 53505.83        & 1728\\
6 & 91414-01-03-01 & 53509.97        & 1936\\
7 & 91414-01-04-00 & 53511.80        & 256\\
8 & 91414-01-04-01 & 53515.17        & 752\\
\hline\end{tabular}
\label{tab:loginteg}
\end{table}

\subsection{\integ\ data reduction}
The \integ\ observations are part of a monitoring campaign on the microquasar \G1915.
Given that the field of views (hereafter FOV) of the main instruments on-board \integ\ are 
rather large (e.g. 29$^\circ\times$29$^\circ$ down to 0 response for IBIS/ISGRI see 
Lebrun et al. 2003), \aql\ is always in the field of view of both IBIS and SPI during 
our campaign. This is, however, not true for the JEM-X monitors that have smaller FOV, as
a consequence, \aql\ is always outside of the JEM-X FOV.\\
\indent The IBIS/ISGRI data were reduced with the version 5.0 of the {\tt Off-line Scientific
Analysis (OSA)} software. The sequence of procedures we performed are similar to those reported
in e.g. Rodriguez et al. (2006), with production of images in the 20--40 and 40--80 keV ranges, 
leaving the software free to find the most significant sources in the field, and extraction of
spectral products for the most active sources in the field. The {\tt OSA 5} response matrices 
were used for spectral analysis. We  extracted spectra from the SPI instrument,
but due to a lower sensitivity than ISGRI below 200 keV, these does not provide additional 
information, and are therefore not included here. 

\subsection{\rxte\ data reduction}
The \rxte\ data were reduced with the version 6.0.2 of the {\tt HEASOFT} software.
For the Proportional Counter Array (PCA), we extracted spectral products from the top layer 
of Proportional Counter Units (PCU) \#0 and \# 2 that were the only ones always turned on 
throughout the outburst. We restrained the spectral analysis of  
the High Energy Timing Experiment (HEXTE) to Cluster A only. For both instruments the data 
reduction was done in a similar manner as  in Rodriguez et al. (2003),  using the 
faint background maps to estimate the PCA background, 
and filtering out time with rate of 
high electron background in PCU 2 (keeping time with {\tt electron2$<$0.1}). 
We then applied 0.6\% systematics to all PCA channels. 
The resultant spectra were fitted simultaneously in {\tt XSPEC v11.3.2}
in the 3--25 keV (PCA) and 18--180 keV (HEXTE) energy ranges. A normalization constant
was included to account for uncertainty in the instrumental cross-calibration. Freezing that of 
the PCA data to 1 the value of the constant for the HEXTE spectra was always found between 0.95 
and 1.1.

\section{Results}
\subsection{\integ\ observations}
\begin{figure*}
\centering
\epsfig{file=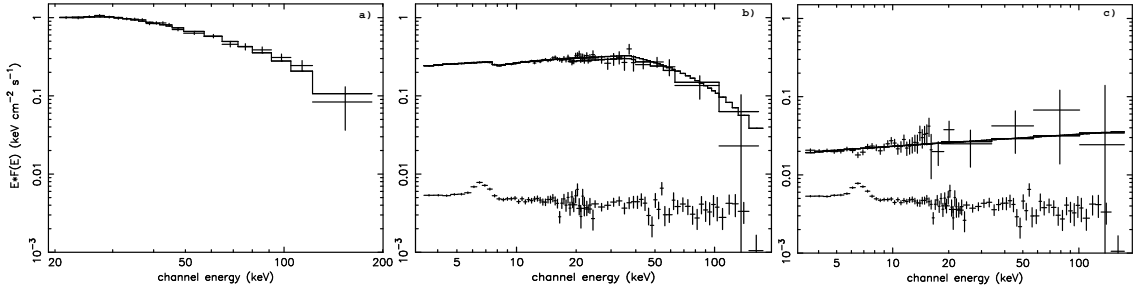,width=15cm}
\caption{a) \integ/ISGRI 20-200 keV unfolded spectrum of Aql X-1 at the maximum of the outburst, during
revolution 305. b\&c) \rxte/PCA and HEXTE spectra of Obs. 1 (b), and Obs. 5 (c). In  panels b \& c 
the lower spectra represent the emission of the Galactic ridge at the position of \aql\ 
as deduced from Revnivtsev (2003).The best fit model is superimposed on the spectrum.}
\label{fig:ISGRI}
\end{figure*}

\aql\ is not detected by \integ/ISGRI during our first observation, with $3 \sigma$ 
upper limits of 0.37 cts/s ($\sim3.3$ mCrab\footnote{Conversions to mCrab are here made 
assuming a Crab-like spectrum.}$\approx2.7\times10^{-11}$ \ergcms) 
and 0.33 cts/s ($\sim5.5$ mCrab$\approx3.9\times10^{-11}$ \ergcms) in the 
20--40 keV and 40--80 keV energy ranges. Our first detection occurs on MJD 53472, 
with fluxes of 16.13 cts/s ($\sim150$ mCrab$\approx1.19\times10^{-09}$ \ergcms) and 
6.7 cts/s ($\sim111$ mCrab$\approx0.79\times10^{-09}$ \ergcms) in the same ranges. 
During our last observation, the flux of \aql\ has significantly decreased to a level 
of 0.8 cts/s (7 mCrab$\approx5.8\times10^{-11}$ \ergcms) in the 20--40 keV range, 
while it is not detected at higher energy with a $3 \sigma$ upper limit of 0.46 cts/s 
($\sim 7.6$ mCrab$\approx5.4\times10^{-11}$ \ergcms) in the 40--80 keV range. The best fit 
spectral parameters are shown in table 2.  The ISGRI spectrum of Rev 305,
 shown in Fig.\ref{fig:ISGRI}, is well described between 20 and 200 keV by a power law with a 
high energy cut-off.  One month later, in Rev 315, the source's hard and soft X-ray flux has 
decreased dramatically and the spectral parameters are poorly constrained. A simultaneous fit 
of the ISGRI Rev 315 and PCA+HEXTE Obs 5 spectra is discussed in section \ref{sec:simult}.

\subsection{\rxte\ observations}
The PCA+HEXTE spectra were first fitted with a simple model of an absorbed power law. We first left
\nh\ free to vary in the fits, but the latter always tended to values close to 0. Given 
the 3 keV lower boundary of the PCA bandpass, we fixed  \nh\  to the Galactic value along 
the line of sight (\nh=0.34$\times10^{22}$ cm$^{-2}$).
A cut-off is statistically required in Obs.1, but is not present in the other 
observations. Obs. 1 spectra were thus fitted with the {\tt cutoffpl}
model in {\tt XSPEC}. In the four first observations, an iron edge is statistically 
required, while no iron emission line seems to be present. In order to limit the number of 
fit parameters we modeled the edge with the simple {\tt edge} multiplicative model. The 
broad band spectrum of Obs. 1 with the best fit model is represented in Fig. \ref{fig:ISGRI}.
In Obs. 4 the value of the iron edge tends to a very low value if it is left completely 
free to vary. We therefore allowed variations only above 6.5 keV. 
 We note that for Obs. 5 onwards the spectrum is very likely to be 
contaminated by the Galactic ridge emission (Fig. \ref{fig:ISGRI} right panel). 
This manifests in particular by the appearance of 
an iron emission line at $\sim6.7$ keV. In order to get rid of this extra component, we modeled
the ridge emission following Revnivtsev (2003), at the position of \aql. The resultant spectra 
were then considered as extra background files for PCA and HEXTE 
(Fig. \ref{fig:ISGRI} right panel). 
After Obs. 5 the source is not significantly detected any more, even when  the 
background emission of the Galaxy is modeled properly.

\begin{table*}
\caption{Best fit parameters obtained from the fits to the \integ/ISGRI and \rxte/PCA+HEXTE spectra, with the model 
of an absorbed (cut-off) power law (either {\tt powerlaw or highecut*powerlaw} when required). 
\chis\ stands for reduced $\chi^2$. All errors are given at the 90\% level. 
Upper limits are given at 90\%. $^\star$The value of the edge was constrained to be higher than 
6.5 keV. The 2-20 keV flux of the \integ\ observations are extrapolated.}
\begin{tabular}{ccccccccc}
\hline
 Obs. & Edge energy & max $\tau$ & $\Gamma$ & E$_{cut}$ & E$_{fold}$& \multicolumn{2}{c}{Unabs. Flux} & \chis \\
        &   (keV)     &            &          &   (keV)        & keV & \multicolumn{2}{c}{($\times 10^{-10}$ \ergcms)} & (DOF) \\
        &              &           &         &                 & & 2-20 keV  & 20-100 keV & \\
\hline
\multicolumn{9}{c}{\integ}\\
Rev. 305     &                       &           &     $1.93\pm0.25$ & $29_{-4}^{+5}$ & $49_{-10}^{+14}$ & 33.9 & 19.0 & 0.98 (19)\\
Rev. 315     &                      &            &     $2.4_{-0.5}^{+0.7}$&           &                  & 4.01 & 1.3 & 0.5 (1)\\            
\hline
\multicolumn{9}{c}{\rxte}\\
Obs. 1 &  7.7$\pm0.2$      &  0.14$\pm0.02$ &  1.88$\pm0.01$ &  37$_{-11}^{+23}$ & $52.9_{-38}^{+45}$& 9.6 & 6.8 & 0.83 (57)\\
Obs. 2 & 7.6$\pm0.2$      &   0.15$\pm0.003$ & 1.91$\pm0.01$ &                    & & 8.2 & 6.8 & 0.96 (59)\\
Obs. 3 & 7.0$\pm0.2$      & 0.14$\pm0.003$ & 1.89$\pm0.02$ &                      & & 5.3 & 4.8 & 1.29 (59)\\
Obs. 4 & $<$7.1$^\star$     & $<$0.18          & 1.93$\pm0.03$ &                    &  & 2.6 & 2.1 & 1.26 (57)\\
Obs. 5 &                  &                & 1.84$\pm0.07$ &                      & & 0.8 & 0.7 & 0.91 (34)\\ 
\hline
\multicolumn{9}{c}{\integ+\rxte}\\
Rev 315./ Obs. 5 & &                   & 1.86$\pm0.05$ &                     & & 0.99 & 0.91 & 0.83 (38)\\
\hline
\end{tabular}
\label{tab:fitpca}
\end{table*}

\subsection{Simultaneous fit of the \integ\ spectrum of Rev. 315 and \rxte\ spectra of Obs. 5}
\label{sec:simult} 
Given the low luminosity of the source at the end of the outburst, and in order to 
try to obtain better constraints on the spectral parameters we fitted the \integ\ and
\rxte\ spectra simmultaneously.
Since the observations are not strictly simultaneous, we needed to choose observations 
that were the closest in terms of spectral state/luminosity.
Table \ref{tab:loginteg} show that Rev. 315 observation is closest 
in time to Obs. 5 of \rxte. In addition the 20--100 keV flux of Rev. 315 is closest to the 
value of the 20--100 keV flux obtained during Obs. 5. We therefore 
fitted the spectrum of Rev. 315 simultaneously with those of Obs. 5. We note that 
No normalization constant is needed, and that a single (absorbed) power law gives 
a good representation of the data. The best fit parameters are reported in Table \ref{tab:fitpca}. 
The spectra from ISGRI and HEXTE are
compatible, giving further confidence in the spectral parameters obtained at the peak of the outburst.

\section{Discussion}
We report the analysis of high energy \integ\ and \rxte\ observations of a faint  
Soft X-ray outburst of \aql\ (F$_{1-12}\sim 90$ mCrab at maximum, Fig. \ref{fig:lc}).
When looking at high energies, however, the situation is quite different. At the peak of 
the outburst the source is clearly detected up to $\sim 150$ keV with \integ\ 
(Fig. \ref{fig:ISGRI}).
Although the lack of simultaneous data 
below 20 keV renders it  difficult to obtain the source photon index with high accuracy, the 
spectrum is typical of a hard state, with a high energy cut-off at $\sim 30$ keV. Assuming this 
model is valid down to 1 keV, the 1-200 keV luminosity is $\sim2\times 10^{37}$ erg/s 
(with $d=5$ kpc).  This 
is the range of luminosities at which \aql\ had
been observed to undergo a transition to a soft state in the 1997 and 1999 outbursts 
(Zhang et al, 1998; Maccarone \& Coppi 2003) (L=$1.6$--$2.2\times 10^{37}$ erg/s 
at 5 kpc in 1999), and clearly much higher than the luminosity at which it had returned 
to the hard state in 1999 ($\sim 2.4\times 10^{36}$ erg/s at 5 kpc, Maccarone 
\& Coppi 2003). 
Note that to check the consistency of the extrapolated flux, 
the web tool {\tt{PIMMS}} was used to estimate the ASM flux given the ISGRI fitted 
spectral parameters, 20-100 keV flux and observed Nh; this returned an ASM count rate 
of $\sim6$ cts/s, which is consistent with the observations in Fig \ref{fig:lc}.
When comparing this episode of hard state with e.g. the initial hard state episode 
of the 1999 outburst (Maccarone \& Coppi 2003), at similar luminosities, we detect the source 
up to higher energy than in 1999. Although this may be the results of the higher sensitivity of 
\integ/ISGRI as compared to \rxte/HEXTE, there are clearly physical differences between both 
episodes. In fact while the luminosity is close to that at which the state transition occured in 1999,
the temperature of the corona (close to the folding energy of the exponential cut-off in 
our model, see e.g. Rodriguez et al. 2003), is higher here ($\sim 50$ keV) than in 1999 
($\sim 10$ keV, Maccarone \& Coppi 2003). This higher temperature of the corona likely 
indicates that the accretion disk is cooler and its inner edge is further out than in 1999. 
A necessary condition for the transition to occur would then be related to physical parameters 
of the disk/corona, and not only to the total luminosity (usually taken to be equivalent to the 
mass accretion rate) of the source. Although based on different behaviors, similar conclusions were drawned 
from  during outbursts of e.g. XTE J1550$-$564 (Homan et al. 2001; Rodriguez et al. 2003).\\
\indent  The \rxte\ monitoring of the source, close to the end of the outburst,
shows that the spectral parameters, in particular the photon power law index, do not 
evolve significantly. The decay corresponds to an overall decrease of the source 
luminosity, compatible with a global decrease of the accretion rate (as seen during 
the outburst decay of transient sources, eg XTE J1550$-$564 (Rodriguez et al. 2003).
{In fact \aql and XTE J1550$-$564 share some interesting similarities: they both underwent 
hard state outbursts (e.g Sturner \& Shrader 2005, and the present study), they both 
showed hard X-ray flares
prior to outbursts (Yu et al. 2003; Yu 2004), the analysis of their power spectra can show 
the presence of  low frequency QPOs at frequencies $1-20$ Hz during periods 
characterized by hard spectra (Yu et al. 2003, Rodriguez et al. 2004b), and they both have 
low period of recurence (although no real period can be claim for either of the sources).}
It is well known that the (normal) outbursts evolution are comparable in both 
the neutron star  and black hole transients  in 
many aspects (e.g. Yu et al. 2003, 2004). We show here that not only normal outbursts are 
comparable, but that both type of systems can undergo failed/mini or low intensity 
hard state outbursts, suggesting that the mechanism driving the transition is similar 
in both sources. During the ``failed outburst'' episode of XTE~J1550$-$564,  Sturner \& Shrader 
(2005) suggested that more than one physical mechanisms could give rise to outbursts in 
this source, and that the faint 
outburst could correspond to a discrete accretion event. This possibility could also 
explain the low state
outburst in Aql~X-1. In that case however the shape of the outburst should be FRED-like
when here it has a shape closer to a triangle (Fig. \ref{fig:lc}). In the future 
a wider spectral coverage during those episodes of hard state outburst in neutron star and 
black hole systems, will help us to understand the parameter(s) triggering or not normal/major
outburst.\\

\begin{acknowledgements}
The authors thank the anonymous referee for his/her fruitful comments.
This work is based on observations with INTEGRAL, an ESA mission with 
instruments
and science data centre funded by ESA member states (especially the PI
countries: Denmark, France, Germany, Italy, Switzerland, Spain), Czech
Republic and Poland, and with the participation of Russia and the USA.
This reserach has made use of data obtained through the High Energy 
Astrophysics Science Archive Center Online Service, provided by the 
NASA/Goddard Space Flight Center.

\end{acknowledgements}

\end{document}